\documentstyle\lbrack 12pt \rbrack{article}
\newcommand{\be}{\begin{equation}}
\newcommand{\ee}{\end{equation}}
\newcommand{\bea}{\begin{eqnarray}}
\newcommand{\eea}{\end{eqnarray}}
\newcommand{\nn}{\nonumber \\}
\newcommand{\e}{{\rm e}}
\newcommand{\pint}{{1 \over 2\pi}\int d^2x\,}

\newcommand{\ffint}{{k \over 4\pi}\int d^2x\,}

\newcommand{\eint}{{k \over 8\pi}\int \sqrt{-g} d^2x\,}
\newcommand{\sqg}{\sqrt{-g}}
\newcommand{\ephi}{\e^{-2\phi}}

\newcommand{\epphi}{\e^{2\phi}}

\newcommand{\papl}{\partial_-}
\newcommand{\pami}{\partial_+}
\newcommand{\papm}{\partial_\pm}
\newcommand{\pamp}{\partial_\mp}

\newcommand{\co}{{\cal O}}
\newcommand{\ie}{{\it i.e.,\ }}

\newcommand{\eg}{{\it e.g.,\ }}

\newcommand{\half}{{1 \over 2}}
\newcommand{\etc}{{\it etc.}\ }

\begin{document}
\begin{titlepage}
\addtolength{\baselineskip}{0.20\baselineskip}
\hfill    NDA-FP-10/93, OCHA-PP-33

\hfill February 1993

\vspace{.5cm}
\begin{center}
\LARGE
{\sc Black Hole from Black Hole in Two Dimensions}

\vspace{.5cm}
\large
Shin'ichi Nojiri

{\it Department of Mathematics and Physics
\\ National Defense Academy
\\ Hashirimizu, Yokosuka 239, JAPAN}

\vspace{.5cm}
\large
Ichiro Oda

{\it Faculty of Science, Department of Physics
\\ Ochanomizu University
\\ 1-1, Otsuka 2, Bunkyo-ku, Tokyo 112, JAPAN}

\end{center}

\vspace{1cm}
\begin{abstract}
We  present  a  new  class  of quantum two dimensional dilaton gravity model,
which is described by $SL(2,R)/U(1)$ gauged Wess-Zumino-Witten model deformed
by $(1,1)$-operator.  We analyze the model by ${1 \over k}$ expansion ($k$ is
the  level  of  $SL(2,R)$  Wess-Zumino-Witten model)  and  we  find  that the
curvature singularity does not appear when $k$ is large and the Bondi mass is
bounded  from  below.  Furthermore,  the rate of the Hawking radiation in the
quantum black  hole  created  by  shock  wave  goes  to  zero  asymptotically
and the radiation stops when the Bondi mass vanishes.
\end{abstract}

\begin{center}
PACS number(s) : 04.60.+n, 11.17.+y, 97.60.Lf
\end{center}

\end{titlepage}
\newpage

Two dimensional dilaton gravity theory proposed by Callan, Giddings, Harvey
and Strominger\cite{I} has been actively studied by many authors
\cite{IA}--\cite{V}.
The consistent quantization of this theory in the conformal gauge
\be
\label{ei}
g_{\mp\pm}=-{1 \over 2}\e^{2\rho}\ ,\ \ g_{\pm\pm}=0 \ ,
\ee
requires that the theory should be a conformal field theory
with the vanishing central charge
and a class of quantum actions was proposed \cite{II,III}.
The actions are given by a sum of the actions of two free fields,
which corresponds to $\rho$ and the dilaton field $\phi$, and
the dilatonic cosmological term which is a $(1,1)$-operator.
This class of theories was, however, shown to have no lower bound in
energy and the Hawking radiation does not go to zero
asymptotically \cite{IV,V},
therefore, this class of theories should be unphysical.

In this paper, we present a new class of quantum dilaton gravity models,
which is described by
$SL(2,R)/U(1)$ gauged Wess-Zumino-Witten model \cite{VI}
deformed by $(1,1)$ operator which corresponds to cosmological term.
We analyze the model by ${1 \over k}$ expansion ($k$ is the level of
$SL(2,R)$ Wess-Zumino-Witten model) and we find that the curvature singularity
does not appear when $k$ is large and the Bondi mass is bounded from below.
Furthermore, the rate of the Hawking radiation in the quantum black hole
created by shock wave goes to zero asymptotically and the radiation stops
when the Bondi mass vanishes.

In the original paper by CGHS \cite{I}, the quantum effects were expected to
be described by adding correction term which only comes from the conformal
anomaly to the classical action.
It was, however, clarified that \cite{II,III,AI}
we need more counterterms since the quantum action should have conformal
symmetry when we choose the conformal gauge (\ref{ei}) \cite{VA}.
By following de Alwis' paper \cite{V}, we assume the kinetic term
of the quantum action in the dilaton gravity coupled to $N$ free bosons
is given by
\bea
\label{eii}
S_{\rm kin}&=&\pint \Bigl\lbrack
-8\e^{-2\phi}(1+h(\phi))\pami\phi \papl\phi \\
&\ &+4\e^{-2\phi}(1+\bar h(\phi))(\pami\phi \papl\rho
+\pami\rho \papl\phi)
+2\kappa(1+\bar{\bar h}(\phi))\pami\rho \papl\rho\Bigr\rbrack\ .
\nonumber
\eea
Here $h(\phi)$, $\bar h(\phi)$ and $\bar{\bar h}(\phi)$ are $\co (\epphi)$
and $\kappa={24-N \over 6}$.
Note that $\epphi$ plays the role of space-time dependent gravitational
coupling \cite{I}.
In Ref.\cite{V}, it was only considered the case $\bar{\bar h}(\phi)=0$,
where $S_{\rm kin}$ can be rewritten by two free fields action.
In this paper, we consider $\bar{\bar h}(\phi)\neq 0$ case.
If we define new fields $X$ and $Y$ by,
\bea
\label{eiii}
X&\equiv&\pm 2b \int^\phi ds \e^{-2s}
\sqrt{(1+\bar h(s))^2(1+\bar{\bar h}(s))^{-1}+\kappa \e^{2s}(1+h(s))} \ , \nn
Y&\equiv&\pm a \Bigl\lbrack\rho
+{2 \over \kappa}\int^\phi ds \e^{-2s}
(1+\bar h(s))(1+\bar{\bar h}(s))^{-1} \Bigr\rbrack\ ,
\eea
the kinetic term of the quantum action $S_{\rm kin}$ in Eq.(\ref{eii})
can be written by,
\be
\label{eiv}
S_{\rm kin}=\ffint \Bigl\lbrack \papl X\pami X
-\Bigl(1+\bar{\bar h}(\phi(X))\Bigr)
\papl Y \pami Y \Bigr\rbrack\ .
\ee
Here $a$ and $b$ are defined by
\be
\label{eiva}
a=\sqrt{-{4\kappa \over k}}\ ,\ \ \ b=\sqrt{-{4 \over k\kappa}}\ ,
\ee
and $k$ is a constant satisfying $k\kappa <0$ and we identify $k$ with the
level of $SL(2,R)$ Wess-Zumino-Witten model later.
If $\bar{\bar h}(\phi)$ in the action (\ref{eiv}) is a constant, this action
describes a free field theory, which is the simplest conformal field theory.
On the other hand, if we can choose $1+\bar{\bar h}(\phi(X))=\tanh^2 X$,
the action (\ref{eiv}) is nothing but the action of another conformal field
theory, \ie $SL(2,R)/U(1)$ gauged Wess-Zumino-Witten model in unitary
gauge \cite{VII}.
\be
\label{eix}
S_{\rm kin}=\ffint \Bigl\lbrack \papl X\pami X
-\tanh^2 X \papl Y \pami Y \Bigr\rbrack\ .
\ee
In fact, if we choose, for example,
\bea
\label{eaii}
1+h(\phi)&=&-{\ephi \over \kappa}
\Bigl\{\Bigl(1-{\kappa\epphi \over 2}\Bigr)^2
\tanh^2(b\ephi)-1\Bigr\} \ , \nn
1+\bar h(\phi)&=&\Bigl(1-{\kappa\epphi \over 2}\Bigr)
\tanh^2(b\ephi) \ , \nn
1+\bar{\bar h}(\phi)&=&\tanh^2(b\ephi) \ ,
\eea
\ie when $X$ and $Y$ are given by,
\be
\label{eaiii}
X=b\ephi\ , \ \ \
Y=a\Bigl(\rho-\phi-{1 \over \kappa}\ephi\Bigr)\ ,
\ee
we find the action (\ref{eiv}) is rewritten in the form of Eq.(\ref{eix}).
Note that there appears non-perturbative contribution of
$\co(\e^{-{C \over \e^{2\phi}}})$, ($C$ is a positive constant).

The cosmological term which is $(1,1)$-operator can be added to
the action (\ref{eix}). The vertex operator $V_{lm}$, which is an $(l,m)$
representation of $SL(2,R)$, is given in unitary gauge by \cite{VIII},
\be
\label{exi}
V_{lm}=(\sinh^2 X)^l\e^{-2mY}F(m-l, -m-l, 1 ; \coth^2 X)\ .
\ee
Here $F(\alpha , \beta , \gamma ; x)$ is Gauss' hypergeometric function.
If we require
$V_{lm}\sim \e^{c(X+Y)} \sim \e^{c'(\rho-\phi)}$ ($c$ and $c'$ are
positive constants.) when $X\longrightarrow \infty$ \ie
$\epphi \longrightarrow 0$ (weak coupling limit),
which is expected from the $\beta$-function analysis \cite{V},
we find $l=- m$ and
\be
\label{exiii}
V_l\equiv V_{ll}=(\sinh^2 X)^l\e^{2lY}\ .
\ee
Then the total action including cosmological term should be given by,
\bea
\label{eixx}
S&=&\ffint \Bigl\lbrack \papl X\pami X
-\tanh^2 X \papl Y \pami Y
+ {\alpha \over 4k}(\sinh X\e^Y)^{2l}\Bigr\rbrack \nn
&\ &+(N\ {\rm free\ boson\ terms})\ .
\eea
When $l$ is complex,  we replace the cosmological term
${\alpha \over 4k}(\sinh X\e^Y)^{2l}$ in Eq. (\ref{eixx}), by
${1 \over 8k}\Bigl\{
\alpha (\sinh X \e^Y)^{2l}+\alpha^* (\sinh X \e^Y)^{2l^*}\Bigr\}$.

In order to consider the stress tensors, we rewrite the action
(\ref{eixx}) in a reparametrization invariant form,
\bea
\label{exx}
S&=&\eint \Bigl\lbrack -g^{\mu\nu} \partial_\mu X(\phi) \partial_\nu X(\phi)
\nn
&\ &+\tanh^2 X g^{\mu\nu}
\partial_\mu \Bigl(a\hat\rho+\hat Y(\phi)\Bigr)
\partial_\nu \Bigl(a\hat\rho+\hat Y(\phi)\Bigr) \nn
&\ &+ {\alpha \over k}\Bigl(\sinh X
\e^{a\hat\rho+\hat Y(\phi)}\Bigr)^{2l}\e^{-2\hat\rho}\Bigr\rbrack
+(N\ {\rm free\ boson\ terms})\ .
\eea
Here we have used the parametrization in Eq.(\ref{eaiii}) and we define
$\hat Y(\phi)$ and $\hat\rho$ by ($R$ is a scalar curvature)
\be
\label{exxi}
\hat Y\equiv a\Bigl(-\phi-{1 \over \kappa}\ephi\Bigr)\ , \ \ \
\hat\rho\equiv -\half (\partial_\mu \sqg g^{\mu\nu} \partial_\nu )^{-1}
\sqg R\ .
\ee
Then we find the stress tensors
$T_{\pm\pm}$ have the following forms,
\bea
\label{exxii}
T_{\pm\pm}&=&k(\papm X \papm X -\tanh^2 X \papm Y \papm Y) \nn
&\ &+{ka \over 2}\papm\int^{x^\mp}dy^{\mp}
\Bigl\{\papl (\tanh^2X\pami Y)+\pami (\tanh^2X\papl Y)\Bigr\} \nn
&\ &-{\alpha \over 4}(1-al)\papm \int^{x^\mp}dy^{\mp}(\sinh X \e^Y)^{2l}
+T_{\pm\pm}^{\rm matter} \ , \nn
T_{\pm\mp}&=&-{ka \over 2}
\Bigl\{\papl (\tanh^2X\pami Y)+\pami (\tanh^2X\papl Y)\Bigr\} \nn
&\ &-{al\over 4}\alpha(\sinh X \e^Y)^{2l} \ .
\eea
Here $T_{\pm\pm}^{\rm matter}$ is the stress tensor of $N$ scalar
fields.
Note that there appear non-local terms in $T_{\pm\pm}$. In order to fix the
boundary condition, we assume that $T_{\pm\pm}$ is given by,
\bea
\label{exxiii}
T_{\pm\pm}&=&k(\papm X \papm X -\tanh^2 X \papm Y \papm Y) \nn
&\ &+{ka \over 2}\papm
\Bigl\{\tanh^2X\papm Y+\int^{x^\mp}_{\alpha^\mp} dy^{\mp}
\papm(\tanh^2X\pamp Y)\Bigr\} \nn
&\ &-{\alpha \over 4}(1-al)\papm
\int^{x^\mp}_{\beta^\mp} dy^{\mp}(\sinh X \e^Y)^{2l}
+T_{\pm\pm}^{\rm matter}
\ .
\eea
The physical quantities like black hole
mass \etc do not depend on the boundaries $\alpha^\pm$ and $\beta^\pm$
if we use the equations of motion.
And as we also discuss later, due to the non-local terms in $T_{\pm\pm}$,
the total central charge  $c^{\rm total}$ of the system and the
dimension $\Delta_l$ of $V_l$ in Eq.(\ref{exiii}) are shifted by $2$ and
$-\half$, respectively,
\bea
\label{exxiv}
c^{\rm total}&=&c^{SL(2,R)/U(1)}+2-26+N \nn
&=&{3k \over k-2}-1+2-26+N \nn
&=&{6 \over k-2}-6\kappa +2 \ , \nn
\Delta_l&=&-{l(l+1) \over k-2}-{l^2 \over k}-\half \ .
\eea
Here $c^{SL(2,R)/U(1)}$ is the central charge of $SL(2,R)/U(1)$
gauged Wess-Zumino -Witten model with level $k$
($c^{SL(2,R)/U(1)}={3k \over k-2}-1$).
Since the total central charge  $c^{\rm total}$ should vanish, we obtain
$\kappa={1 \over 3}+{1 \over k-2}$.
Since $k\kappa<0$, we find $k<-1$ or $0<k<2$ \ie we find the
restriction for the number of matter fields $N$ \ie $22<N<24$ or $25<N$.
In the present formulation, the plausible approximation is made when the
absolute value of $k$ is large, so that we have to restrict ourselves to
the case $k<-1$, in other words, $22<N<24$. More general formulation
will be reported in the future publication \cite{IX}.

We now solve the equations of motion and constraints of the system.
In order to do this, it is convenient to define new fields $X^\pm$ by,
\be
\label{exxvi}
X^\pm=\pm\sinh X\e^{\pm Y} \ .
\ee
Then the action (\ref{eixx}) is rewritten by
\bea
\label{exxvii}
S&=&\ffint \Bigl\lbrack -{\papl X^+ \pami X^- +\papl X^- \pami X^+
\over 2(1-X^+ X^-)}
+ {\alpha \over 4k}(X^+)^{2l}\Bigr\rbrack \nn
&\ & +(N\ {\rm free\ boson\ terms})\ .
\eea
and we obtain the following equations of motion
\bea
\label{exxviia}
0&=&{X^-\papl X^+ \pami X^+ \over (1-X^+ X^-)^2}
+ {\papl \pami X^+ \over 1-X^+ X^-} \ , \\
\label{exxviib}
0&=&{X^+\papl X^- \pami X^- \over (1-X^+ X^-)^2}
+ {\papl \pami X^- \over 1-X^+ X^-}
+ {l\alpha \over 2k}(X^+)^{2l-1}
\ .
\eea
Note that $X^+=w$ ($w$ : constant) satisfies the first equation
(\ref{exxviia}). Then we can solve the second equation (\ref{exxviib}),
\be
\label{exxix}
X^+=w\ , \ \ \
X^-={1 \over w}\Bigl(1-\e^{-\Lambda x^+x^-+u(x^+,x^-)}\Bigr) \ .
\ee
Here $\Lambda\equiv -{l\alpha \over 2k}(w)^{2l}$ and
$u(x^+,x^-)=u_+(x^+)+u_-(x^-)$.
It is important that this class of special solutions includes the solutions
corresponding to the classical black hole and the linear dilaton vacuum
in the weak coupling region.
{}From now on, we only consider the solutions given by Eq.(\ref{exxix}).
By using the solution (\ref{exxix}), we find the stress tensors
in Eqs.(\ref{exxii}), (\ref{exxiii}) have the following forms,
\be
\label{exxxi}
T_{\pm\mp}=0 \ , \ \ \ T_{\pm\pm}=
-{ka \over 4}\partial_{\pm}^2u_\pm(x^\pm)
+T_{\pm\pm}^{\rm matter} \ .
\ee
Now let us impose the only a priori restriction \cite{V}
in the present formalism given by,
\be
\label{exxxii}
0=T_{\pm\pm}+t_{\pm\pm} \ , \ \
t_{\pm\pm}=-{26 \over 24}{1 \over (x^\pm)^2} \ .
\ee
When $T_{\pm\pm}^{\rm matter}=0$, we find the following  solutions
\be
\label{exxxiii}
-{ka \over 4}u_\pm(x^\pm)
=a_\pm+b_\pm x^\pm -{26 \over 24}\ln |x^\pm| \ .
\ee
If $b_+=b_-=0$, the solution (\ref{exxxiii}) corresponds to a static object.
On the other hand, the solution corresponding to the shock wave which
describes collapsing matters,
$T_{\pm\pm}^{\rm matter}=m\delta(x^+-x^+_0)$, is given by
\bea
\label{exxxv}
-{ka \over 4}u_+(x^+)
&=&a_++b_+ x^+ -m(x^+-x^+_0)\theta(x^+-x^+_0)
-{26 \over 24}\ln |x^+| \ , \nn
-{ka \over 4}u_-(x^-)
&=&a_-+b_- x^- -{26 \over 24}\ln |x^-| \ .
\eea

We now discuss the Bondi mass for the solutions (\ref{exxxiii}) and
(\ref{exxxv}).
Since the action (\ref{eixx}) is correct for large $k$ \cite{VII},
we consider the large $k$ limit ($k \longrightarrow -\infty$,
$\kappa \longrightarrow {1 \over 3})$.
The discussions which are not based on ${1 \over k}$
expansion will be given elsewhere \cite{IX}
by using the action in Landau gauge \cite{VIII}.

By following de Alwis \cite{V}, we consider the first variation of the stress
tensor $\delta T_{\pm\pm}$ around some reference solution to which a solution
asymptotically approaches at future null infinity and we define the Bondi mass
as follows,
\be
\label{exxxvi}
M(\bar x^-)=-\int^{{\cal I}^+_R} d\bar x^+ (\delta T_{++}+\delta T_{+-})
\ .
\ee
Here $\bar x^\pm$ are the asymptotically Minkowski coordinates
and ${\cal I}^+_R$ is the future null infinity line.
When we choose the reference solution where \eg
$-{ka \over 4}u=-{26 \over 24}\ln | x^+x^-|$, which is a quantum analogue of
the linear dilaton vacuum in the classical theory,
and the asymptotically Minkowski coordinates
$\bar x^\pm$ by
\be
\label{exxxix}
\bar x^\pm=\pm{1 \over \lambda}\ln(\pm \lambda x^\pm) \ .
\ee
the Bondi mass is given by,
\bea
\label{eaai}
M(\bar x^-)&=&w\sqrt{-k\kappa}\Bigl\{-\lambda+(\pami - \papl)
+\co ((-k)^{-\half})\Bigr\} \nn
&\ &\times\Bigl\lbrack
\e^{{\lambda a \over 2}(\bar x^+-\bar x^-)-{4 \over ka}\Bigl\{
-\e{\lambda(\bar x^+-\bar x^-)}+{26 \over 24}\lambda (\bar x^+-\bar x^-)
-{26 \over 24}\ln \lambda^2\Bigr\}} \delta X^-\Bigr\rbrack
\Bigl|_{{\cal I}^+_R} \ .
\eea
Here $\lambda^2\equiv \Lambda\sqrt{-{k\kappa \over 4}}$ ($\lambda>0$)
and we have only considered the variation of $\delta X^-$ and fixed
$X^+$ since $w$ in Eq.(\ref{exxix}) can be always absorbed into the
redefinition of the coordinates.
Furthermore we have assumed $Y$ is transformed by the coodinate transformation
(\ref{exxxix}) as $Y(x^\pm) \longrightarrow Y(\bar x^\pm)
+{\lambda a \over 2}(\bar x^+-\bar x^-)$
since conformal mode $\rho$ transforms as
$\rho(x^\pm) \longrightarrow \rho(\bar x^\pm)
+{\lambda \over 2}(\bar x^+-\bar x^-)$.
Then we find the Bondi mass for the static solution ($b_+=b_-=0$) in
Eq.(\ref{exxxiii}),
\be
\label{exxxvii}
M(\bar x^-)=\lambda\sqrt{-k\kappa}\Bigl\{
\e^{-{4 \over ak}(a_+ +a_-)}-1\Bigr\}\ .
\ee
Note that the Bondi mass (\ref{exxxvii}) is apparently bounded from below,
which should be contrasted with the results in Ref.\cite{IV,V}.

We now consider the dynamic solution (\ref{exxxv}) when $a_\pm=b_\pm=0$,
which corresponds to the classical solution where matter collapsing
into a linear dilaton vacuum creats a black hole.
Here we choose the asymptotically Minkowski coordinates
$\bar x^\pm$ in Eq.(\ref{eaai}) by, instead of Eq.(\ref{exxxix})
\be
\label{exxxxi}
\bar x^+={1 \over \lambda}\ln(\lambda x^+) \ , \ \ \
\bar x^-=-{1 \over \lambda}\ln(-\lambda x^- -{m \over \lambda}) \ .
\ee
Then the Bondi mass is given by,
\bea
\label{exxxxii}
M(\bar x^-)&=&\lambda\sqrt{-k\kappa}\Bigl\lbrack -1 \nn
&\ &+\Bigl\{1 +{4 \over ka}{26 \over 24}
{m\e^{\lambda \bar x^-} \over \lambda + m\e^{\lambda \bar x^-}}\Bigr\}
\e^{-{4 \over ka}\Bigl\{{m \over \lambda}
\e^{\lambda \bar x^+_0}-{26 \over 24}\ln \Bigl(1+{m \over \lambda}
\e^{\lambda \bar x^-}\Bigr)\Bigr\}} \Bigr\rbrack
\ ,
\eea
where $\bar x^+_0\equiv {1 \over \lambda}\ln(\lambda x^+_0)$.
In the limit of infinite past ($\bar x^- \longrightarrow -\infty$)
in the light-cone frame, the Bondi mass $M(\bar x^-)$ approaches to a constant
$M(\bar x^-)\longrightarrow \lambda\sqrt{-k\kappa}\Bigl\lbrack -1
+\e^{-{4 \over ka}{m \over \lambda}
\e^{\lambda \bar x^+_0}} \Bigr\rbrack$. On the other hand,
in the limit of infinite future ($\bar x^- \longrightarrow +\infty$),
the rate of the Hawking radiation ${d M(\bar x^-) \over d x^-}$ goes to
vanish and the Bondi mass $M(\bar x^-)$ approaches to a constant
$M(\bar x^-)\longrightarrow -\lambda\sqrt{-k\kappa}$.
The constant $-\lambda\sqrt{-k\kappa}$ could be interpreted
as the vacuum energy.

The above results suggest that the curvature singularity does not appear
when $|k|$ is large enough.
If we rewrite $X^-$ in Eq.(\ref{exxix}) as
$X^-={1 \over w}\Bigl(1-\e^{f(x^+,x^-)}\Bigr)$,
the scalar curvature $R\sim \e^{-2 \rho}\papl \pami \rho$ is given by
\be
\label{eaaii}
R\sim f^{{1 \over a}+1}\Bigl\{{\papl\pami f \over f}
-{\papl f \pami f \over f^2}\Bigr\} \ ,
\ee
when $f\longrightarrow 0$. Therefore $R$ diverges in general. However,
since ${1 \over a}$ is $\co ((-k)^\half)$ positive quantity
(see Eq.(\ref{eiva})),
there does not appear curvature singularity if we choose $-k$ large enough.
Since our analysis is correct when $-k$ is large enough, we expect
that no curvature singularity appears for any $k$.

Since the action (\ref{eixx}) is correct for large $k$ \cite{VII},
it is convenient to consider the corresponding action in Landau gauge in
order to clarify what kind of conformal theory describes the system.
The Landau gauge action is given by
\bea
\label{exxxxiii}
S&=&\ffint \Bigl\lbrack \papl X\pami X
-\cosh^2 X \papl Z \pami Z -\sinh^2 X \papl Y \pami Y \nn
&\ &-(\cosh^2X-\half)\Bigl(\papl Z \pami Y - \papl Y \pami Z)
+{1 \over 4}\papl \varphi \pami \varphi \nn
&\ & +{\alpha \over 4k}(\sinh X\e^{Y+{\varphi \over 2}})^{2l}\Bigr\rbrack
+({\rm ghost\ terms}) \nn
&\ &+(N\ {\rm free\ boson\ terms})\ .
\eea
Here $Z$ is a degree of freedom in $SL(2,R)$ Wess-Zumino-Witten model
which is independent of $X$ and $Y$.
A real scalar field $\varphi$ comes from the gauge fields.
By using new fields $T$ and $S$, we can
rewrite the action (\ref{exxxxiii}) in a reparametrization invariant
and local form,
\bea
\label{exxxxiv}
S&=& {k \over 8\pi}\int d^2x\,\Bigl\lbrack
\sqrt{-g} \Bigl\{
-g^{\mu\nu} \partial_\mu X(\phi) \partial_\nu X(\phi)
-\cosh^2 X g^{\mu\nu} \partial_\mu Z \partial_\nu Z \nn
&\ &+\sinh^2 X g^{\mu\nu}
\partial_\mu \Bigl(aT+\hat Y(\phi)\Bigr)
\partial_\nu \Bigl(aT+\hat Y(\phi)\Bigr) \nn
&\ &+ {\alpha \over 4k}\Bigl(\sinh X
\e^{aT+\hat Y(\phi)}\Bigr)^{2l}
\e^{-2T} +{1 \over 4}g^{\mu\nu} \partial_\mu \varphi \partial_\nu \varphi\nn
&\ &+\half g^{\mu\nu} \partial_\mu S \partial_\nu T
+{1 \over 4}SR \Bigr\}-(\cosh^2X-\half) \epsilon^{\mu\nu}\partial_\mu Z
\partial
_\nu
\Bigl(aT+\hat Y(\phi)\Bigr)
\Bigr\rbrack\nn
&\ &+(U(1)\ {\rm ghost\ term})+(N\ {\rm free\ boson\ terms})
\ .
\eea
Here $\hat Y(\phi)$ is defined by Eq.(\ref{exxi}).
By using the equation of motin which we obtain by the variation of $S$,
we find
$T=\half (\partial_\mu \sqg g^{\mu\nu} \partial_\nu )^{-1}
\sqg R$.
In the conformal gauge, this action (\ref{exxxxiv}) has the following form
\bea
\label{exxxxivb}
S&=&\ffint \Bigl\lbrack \papl X\pami X
-\cosh^2 X \papl Z \pami Z -\sinh^2 X \papl \bar Y \pami \bar Y \nn
&\ &+(\cosh^2X-\half)\Bigl(\papl Z \pami \bar Y - \papl \bar Y \pami Z)
+{1 \over 4}\papl \varphi \pami \varphi \nn
&\ &+ {\alpha \over 4k}(\sinh X\e^{-\bar Y+{\varphi \over 2}})^{2l}
\e^{-2\bar T} -{1 \over 4}(\papl S \pami \bar T + \pami S \papl T)
\Bigr\rbrack\nn
&\ & +({\rm ghost\ terms})+(N\ {\rm free\ boson\ terms}) \ .
\eea
Here we have redefined new fields $\bar Y$ and $\bar T$ by
$\bar Y\equiv\sqrt{-{4\kappa \over k}}T+\hat Y(\phi)$ and
$\bar T\equiv T-\rho$.
Then we find that the stress tensors are given by
\bea
\label{exxxxv}
T_{\pm\pm}&=&k\Bigl(\papm X \papm X -\cosh^2 X \papm Z \papm Z
+\sinh^2 X \papm \bar Y \papm \bar Y \nn
&\ &+ {1 \over 4}\papm \varphi \papm \varphi
+\half\papm S \papm \bar T +{1 \over 4}\papm^2 S \Bigr)
+T_{\pm\pm}^{\rm matter} \ , \nn
T_{\pm\mp}&=&{al\over 4}\alpha
(\sinh X \e^{\bar Y+{\varphi \over 2}})^{2l}\e^{-2\bar T}
-{k \over 4}\papl \pami S \ .
\eea
When $\alpha=0$, the system is a direct product of those of $SL(2,R)$
Wess-Zumino-Witten model, free boson $\varphi$, ghosts, $N$ free bosons
and $S$-$\bar T$ fields which contribute to the central charge by $2$.
Since the total central charge $c^{\rm total}$ does not depend on $\alpha$,
we find the central charge $c^{\rm total}$ in Eq.(\ref{exxiv}).
The operator $\bar V_l$ which corresponds to $V_l$ in Eq.(\ref{exiii})
is given by,
\be
\label{exxxxvi}
\bar V_l=(\sinh X\e^{\bar Y+{\varphi \over 2}})^{2l}\e^{-2\bar T} \ .
\ee
Since $\e^{-2\bar T}$ has the conformal dimension $\half$,
$V_l$ has the dimension $\Delta_l$ in Eq.(\ref{exxiv}).
The detailed discussion of the Hawking radiation \etc by using Landau gauge
action would be reported elsewhere \cite{IX}.

In the recent paper \cite{AVII}, the dilatonic supergravity has been
discussed and it has been found classical solution which describes black
hole or linear dilaton vacuum.
It was, however, claimed that there does not appear any solution
corresponding classical black hole in the quantum theory,
which is described by free supersymmetric action deformed by
$(\half,\half)$-operator.
We can now consider new class of quantum action by supersymmetrizing the
action (\ref{eixx}).
By integrating auxiliary fields, we can easily find that the cosmological
term vanishes and therefore there is any solution corresponding to
the classical black hole or the linear dilaton vacuum.

The authors would like to acknowledge H. Kawai, A. Sugamoto and K. Odaka
for valuable discussions.
One of the authors (I.O.) is financially supported by the Japan Society for
the Promotion of Science.


\pagebreak

\end{document}